\documentclass[%
 reprint,
 superscriptaddress,
groupedaddress,
 amsmath,amssymb,
 aps,
 prl,
]{revtex4-2}

\usepackage{graphicx}
\usepackage{soul}
\usepackage{mathrsfs}%
\usepackage{gensymb}
\usepackage{comment}
\usepackage{float}
\usepackage{color}

\newcommand{\new}[1]{\textcolor{black}{#1}}
\usepackage{graphicx}
\usepackage{dcolumn}
\usepackage{bm}
\usepackage{hyperref}
\usepackage[mathlines]{lineno}

\begin{abstract}
Leaves shed by deciduous trees contain 40\% of the annually sequestered carbon, and include nutrients vital to the expansion and health of forest ecosystems. To achieve this, leaves must fall quickly to land near the parent tree -- otherwise, they are lost to the wind, like pollen or gliding seeds. However, the link between leaf shape and sedimentation speed remains unclear. To gauge the relative performance of extant leaves, we developed an automated sedimentation apparatus (ASAP) capable of performing $\sim100$ free fall experiments per day on biomimetic paper leaves. The majority of 25 representative leaves settle at rates similar to our control (a circular disc). Strikingly, the \emph{Arabidopsis} mutant \emph{asymmetric leaves1 (as1)} fell $15\%$ slower than the wild type. Applying the \emph{as1}-digital mutation to deciduous tree leaves revealed a similar speed reduction. Data correlating shape and settling across a broad range of natural, mutated, and artificial leaves support the \emph{fast-leaf-hypothesis}: Deciduous leaves are symmetric and relatively unlobed in part because this maximizes their settling speed and concomitant nutrient retention. 
\end{abstract}
\begin{document}

\title{Settling aerodynamics is a driver of symmetry in deciduous tree leaves}
\author{Matthew D. Biviano}

\affiliation{Department of Physics, Technical University of Denmark}
\author{Kaare H. Jensen}
\email{khjensen@fysik.dtu.dk}
\affiliation{Department of Physics, Technical University of Denmark}

\date{\today}%
\maketitle
\begin{figure*}%
\centering
\includegraphics[width=1\textwidth]{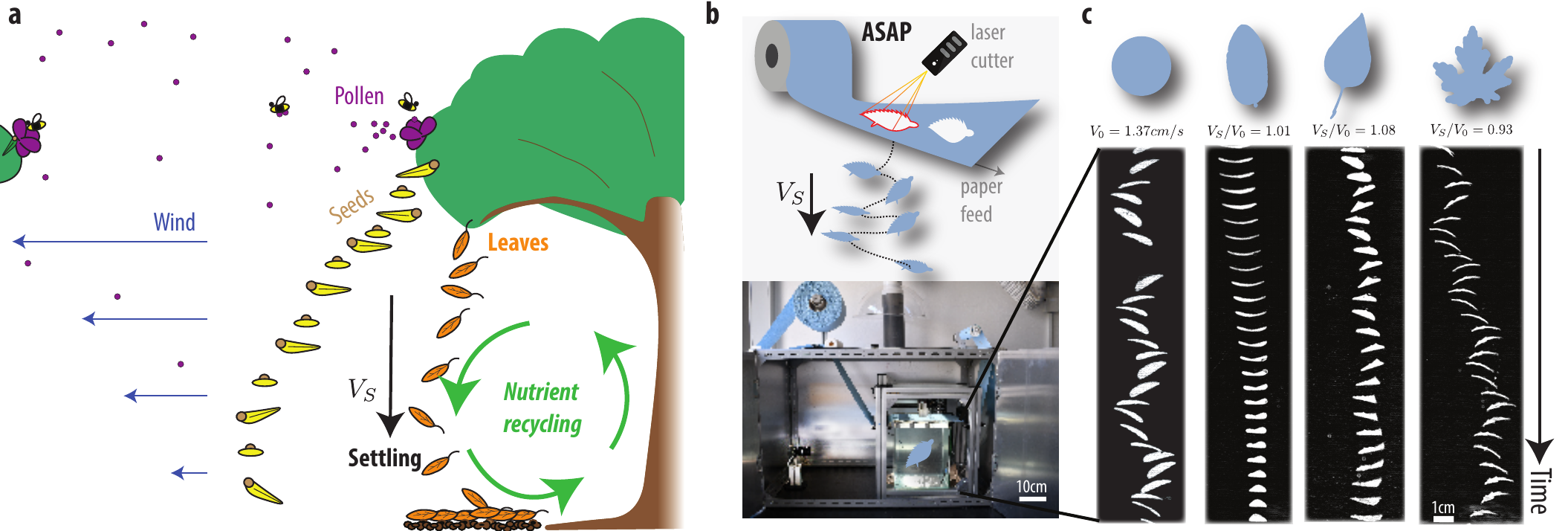}
\caption{The fast-leaf hypothesis. (a) Trees shed pollen and seeds which are dispersed aided by the wind (and by animals). Leaves shed by deciduous trees include nutrients vital to the expansion and health of forest ecosystems. However, the connection between the sedimentary behavior of the leaves and their shape remains unexplored. We propose that the fate of leaves serves the role of enriching the local environment, thus evolving to symmetric shapes which maximise the settling speed $V_S$. (b) Schematic of the Automated Sedimentation Apparatus (ASAP), and a photograph of the device. (c) Representative sedimentation data showing from left to right our control (a circular disc), \textit{Amanchelier arborea}, \textit{Syringa reticulata}, and \textit{Acer saccharinum}. See also video 1 (Fig. \ref{Fig13}).}\label{fig1}
\end{figure*}

Deciduous trees annually shed many parts of themselves; including leaves, pollen, seeds, and fruits (Fig. \ref{fig1}a). Seeds and pollen have evolved aerodynamic traits allowing them to be dispersed far and wide assisted by the wind  \cite{tackenberg2003assessment,de2008effects}. However, despite the ubiquity of falling leaves, it is unclear how their aerodynamic properties interact with the evolutionary story of the tree. In this work, we propose that symmetry in deciduous leaf shape is evolutionarily driven by an aspiration to fall rapidly. This in turn facilitates local nutrient recycling through the soil, thus promoting the fitness of trees and their offspring.

Deciduous trees shed their leaves annually, losing approximately $\sim40\%$  of their assimilated carbon, and also substantial quantities of nutrients (e.g., N, P, K, Mg, Ca) \cite{anderson2021carbon}. Evolutionary thinking suggests that trees may optimise their leaves to deposit where their nutrients can be recycled, and locally improve soil conditions  \cite{sayer2006leaflitternutrition}: 
\new{species that use limiting nutrients more efficiently should have a competitive advantage \cite{hobbie2015plant}}. Natural selection has promoted distinct physical traits that enhance the dispersal of, e.g., seeds, fruits, pollen, across a relatively broad geographic range  \cite{tackenberg2003assessment}. Shape adaptations that take advantage of aerodynamic effects allow, e.g., the parachuting dandelion seed, the gliding Javan cucumber seed, or the helicopter maple seeds to fall slowly and be carried by the wind to settle remarkably far from the parent plant \cite{viola2022flying,bozdag2023novo,rabault2019curving}. In contrast to pollen and seeds, however, leaves must  fall to the ground quickly to achieve a proximal concentration of nutrients. Indeed, leaf removal from \textit{Fagus sylvatica} groves was shown to remove 1.5t/ha of carbon from the soil, corresponding to 30-70\% of the vital nutrients and with concomitant effects on growth \cite{sayer2006leaflitternutrition}. \new{Other potentially beneficial effects of leaf anti-dispersal include shading of competing species \cite{facelli1991indirect}, mediating earthworm activity \cite{kooijman2019litter}, and influencing moss growth via the soil chemical composition \cite{natalia2008effects}, }


The starkest difference between leaf and seed pod aerodynamic design is the relatively high symmetry of leaves. 
The molecular origins of symmetry during leaf morphogenesis are well-established (e.g., \cite{byrne2000asymmetric}), however, \new{exploring the quantitative} reason\new{s} \emph{why} leaves are symmetric  \new{has attracted relatively little attention}. 

The evolutionary origins of extant leaf shape has been placed into the context of, e.g., developmental constraints, reconfiguration to minimize wind loads, gas exchange, and temperature control (e.g., \cite{de2008effects,vogel2012life,vogel2009,de2008effects}). 
\new{Biomechanical rationales for symmetric transport networks have also been proposed, including resilience to damage \cite{katifori2010damage} and mechanical stability \cite{ronellenfitsch2021optimal}. Several authors have advocated for symmetry as a basal state in leaves \cite{damerval2017evolution,cronk2004developmental}.  There is some experimental evidence to support this idea. For instance, some early multicellular organisms (e.g., red algae) comprised approximately symmetric cell clusters \cite{bengtson2017three}, and similar patterns have been described in early land plants \cite{harrison2018origin}. However, several extant species have asymmetric leaves; for example in genii such as Eucalyptus \cite{migacz2018comparative} and Begonia \cite{permata2022morphological}. Similarly, bilateral symmetry is not unambiguous in the fossil record:  asymmetry is present in green algae \cite{tang2020one} and \emph{A. mackiei} leaves \cite{hetherington2021evidence,kerp2018organs}.}

\new{
The available evidence does not allow us to rule out the possibility that symmetry is not a basal trait. 
}
In particular, it remains unclear how leaf \new{shape} interacts with the settling and subsequent nutrient recycling processes.  
Here, we examine the idea that rapid settling associated with the leaf recycling process limits the diversity of leaf shapes. Exploring representative mutations to existing leaf morphologies, we demonstrate that rapid settling explains the ubiquity of symmetric blades.




\section{Results}
Each leaf has its own unique shape and falls in a distinct way depending on factors such as growth patterns and atmospheric conditions \cite{peppe2011sensitivity}. To ensure accuracy and prevent any unwanted confounding effects, we developed a standardized experimental setup to isolate leaf shape effects (Fig. \ref{fig1}b): The Automated Sedimentation Apparatus (ASAP) enables us to examine the falling trajectory of settling paper leaves. Shapes cut by a laser from a ticker tape paper roll fall directly into our observation tank, where a camera records the leaf motion (Fig. \ref{fig1}c). This vertical configuration obviates the need for multi-axis robotic manipulation and vacuum lines in previous studies \cite{howison2020large}.
To isolate the shape as a variable, we maintained a constant leaf blade area $A=100$ mm$^2$. \new{We observed no significant difference in the settling velocity when varying the leaf mimic's absolute area (Fig. \ref{FigS5}) or initial orientation (Fig. \ref{FigS6}).} 
For further information on the tracking software and setup, please refer to the Methods section.

\new{Leaves falling freely in air experience Reynolds numbers ($Re$) ranging from a few hundred to a few thousand, while leaves fixed to the plant body can reach $Re=10^4$ in windy conditions \cite{de2008effects}.}
\new{Our experiment operated in the range}  $Re = v_sD/\nu\approx 100-300$, where $D\approx 1$ cm is leaf radius, $v_s\approx 1$ cm/s is the typical settling speed, and $\nu = 10^{-6}$ m$^2$/s is the viscosity of our operating fluid (water). \new{Consistent with prior literature on drag of freely falling non-spherical particles \cite{bagheri2016drag}, the relative performance different shapes in our experiment did not depend strongly on $Re$ (Fig. \ref{FigS9})}.

Like in a natural leaf settling experiment, the kinematics are highly dependent on initial conditions. Each shape trajectory was therefore recorded $20-30$ times to determine the mean settling speed $V_S$ to within 3\% relative error. The relative settling speeds reported here are thus representative of the many thousands of leaves lost by each tree. By observing the relationship between leaf shape, velocity, and falling behavior across various shapes (Fig. \ref{fig1}c, Video 1), we can gain insights into aerodynamics, drag, and leaf performance optimization.

\begin{figure*}%
\centering
\includegraphics[width=1\textwidth]{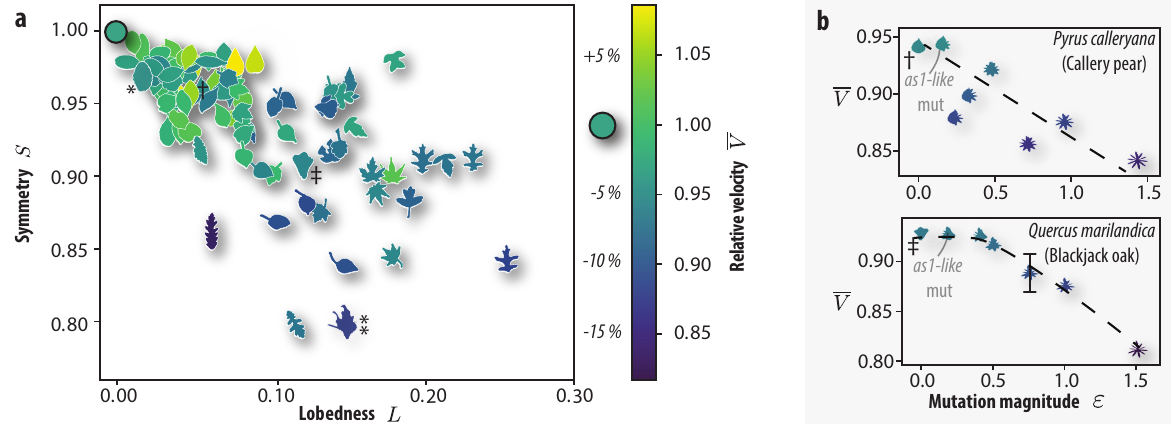}
\caption{Exploring the interplay between leaf shape and terminal sedimentation speed using digital mutations. (a) Leaf morphology quantified by symmetry $S$ and lobedness $L$ (Fig. \ref{Supps2}) and their effect on terminal settling velocity. The terminal velocity (color bar) is scaled by the terminal velocity of the circular control disc. \textit{Arabidopsis thaliana}, wild type (*) and \textit{as1} mutant leaves (**) are highlighted, as well as pear and oak leaves (\dag \ and \ddag) which were subjected to digital mutations. (b) The effects of digital mutation on terminal velocity. Pear and oak leaves were progressively mutated by adding a short wavelength sinusoid to segments of their boundary. We quantify the digital mutation magnitude $\varepsilon=\frac{\Delta A}{A_0}$, where $A_0$ is the area of the original shape, and $\Delta A$ is the area modification of the shape. An error bar is placed in the \textit{Quercus marilandica} dataset to demonstrate the typical variance of the mean sedimentation velocity for that shape. Dashed lines are provided as a guide the eye to highlight the main data trend: Settling speed decreased with increasing mutation amplitude.
}\label{fig2}
\end{figure*}

We investigated the sedimentation characteristics of $25$ species from the families \new{Ulmaceae, Styracaceae, Malvaceae, Sapindaceae, Rosaceae, Fagaceae, Bignoniaceae, Oleaceae, Ebenaceae, Eucommiaceae, Moraceae, Magnoliaceae, Betulaceae, Platanaceae, and Salicaceae} (Fig. \ref{Supps1}).  This dataset \new{comprised deciduous trees with simple leaves that grew to a sufficient height at maturity ($>$5m tall), and contained examples of }major patterns that may affect the aerodynamic performance such as symmetry, lobedness, aspect ratio, toothedness and combinations of those parameters \cite{coombes2014book}. \new{The annual herb \emph{Arabidopsis thaliana} was also included because the \textit{as1}-mutant presents asymmetry \cite{byrne2000asymmetric}. Three leaves from each species were analysed}.

To classify the shapes, we assess the leaves by their reflection symmetry \cite{stoddard2017avian} and their lobedness (Fig. \ref{Supps2}). The leaf symmetry number $S$ quantifies reflection symmetries ($S\ll1$ no symmetry, $S=1$ one or more axes), while the lobedness number $L$ quantifies the mass distribution ($L\ll1$ unlobed, $L=1$ highly lobed). 


Deciduous leaves tend to be symmetric ($S=0.85-1$) and subtly lobed ($L=0-0.2$), see Fig. \ref{fig2}. For instance, \textit{Arabidopsis}  is symmetric and unlobed $(L,S)=(0.02,0.96)$, while some members of the \textit{Quercus} and \textit{Acer} families are lobed and somewhat asymmetric $(L,S)=(0.2,0.9)$. Despite this variation, most extant leaf shapes fall at a velocity within $\pm10$\% of a circular disk of equal mass. 
Interestingly, \textit{Arabidopsis} \textit{as1} mutants leaves settle notably slower ($-15\%$). In this mutant, organ patterning is disrupted and the leaf phenotype is asymmetric (Fig. \ref{fig2}) $(L,S)=(0.14,0.80)$. This leads us to question whether a causal link between symmetry, lobedness, and settling speed exists.


To tease out the effect of a leaf's shape on it's aerodynamics, we applied a shape modification to extant leaf outlines. Based on the aforementioned \textit{as1}'s characteristic large amplitude/short wavelength perturbations, we create \emph{digital mutations} by sinusoidally perturbing segments of the leaf edge. This allows us to gauge the relative performance of different phenotypes that are, in principle, immediately accessible to natural selection. 
We applied {digital mutations} to the bell-shaped Blackjack oak (\textit{Quercus marliandica}) and the ovular Callery pear (\textit{Pyrus calleryana})  (Fig. \ref{fig2}b) leaves. As the mutation strength (modified area fraction $\varepsilon$) grew, a striking $15$\% drop in the terminal velocity was observed in both cases. The pear leaf responded noticably to even small/asymmetric pertubations while the oak's initial responds was less pronounced. 


\begin{figure*}
\centering
\includegraphics[width=1\textwidth]{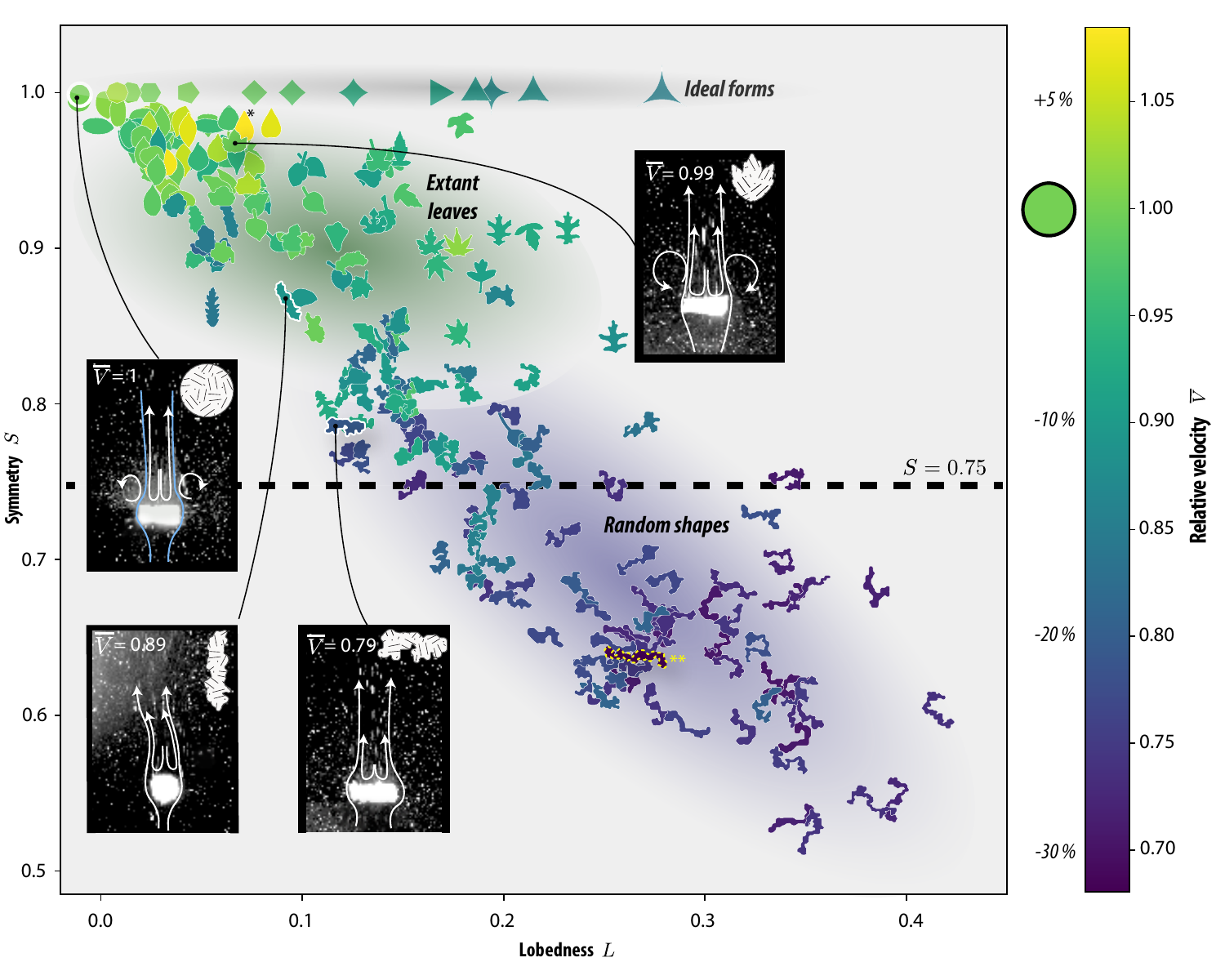}
\caption{The effect of symmetry and lobedness on the terminal velocity. The terminal velocity of 25 extant leaves, 10 ideal forms, and 120 shapes generated from random walk outlines (Fig. \ref{Supps3}) are compared.  High-symmetry shapes ($S>0.75$, dashed horizontal line) settled at speeds on par with a circular disc (within $\pm 10 \%$), and overall the speed decreased with increasing lobedness in this regime. In contrast, the settling of asymmetric objects ($S<0.75$) was substantially slower: speeds systematically decreased $-15\%$ to $-30\%$, and was largely insensitive to the shape's lobedness. The fastest, \textit{Syringa reticulata} (+10\%, *), and slowest, a slender meandering form  (-30\%,**), are labeled for reference. To probe the fluid flow behaviors of various shapes, a laser sheet and 0.02 mm tracer particles dispersed in the solution were used to generate the inset images (see magnified view in Fig. \ref{FigS8}). 
}\label{fig3}
\end{figure*}

Having explored the shape and settling characteristics of natural and mutated leaves, we can now formulate the \emph{fast-leaf-hypothesis}: natural leaves are highly symmetric and relatively unlobed because this maximizes their settling speed and the concomitant nutrient retention. The aforementioned data support this conjecture; however, important questions remain unanswered. Specifically, we inquire how plant leaf shape and sedimentation traits are positioned among all possible 2D geometries? To further explore the impact of symmetry and lobes on sedimentation, we generated 120 random shapes from the outline of the path traced by diffusing particle (\cite{hunt2007relative},  Fig. \ref{fig3}, Fig. S3). The set of 2D forms now covered a relatively wide qualitative and quantitative range of morphologies: From regular polygons across leaves to random structures ($L=0-0.4$ and $S=0.5-1$).

Sedimentation experiments on this broad set of shapes revealed two regions, roughly separated by a symmetry value of $S=0.75$ (Fig. \ref{fig3}). High-symmetry shapes ($S>0.75$) settled at speeds on par with a circular disc (within $\pm 10 \%$), and overall the speed decreased with increasing lobedness in this regime. 
In contrast, the settling of asymmetric objects ($S<0.75$) was substantially slower: speeds systematically decreased by  $-15\%$ to $-30\%$, and was largely insensitive to the shape's lobedness. \new{Generally speaking, asymmetric leaf mimics rotated slowly while falling steadily (Fig. \ref{FigS7}). We speculate that irregular morphologies reduced flow features often associated with fluttering and tumbling. Indeed, } the stark reduction in speed from high to low symmetry also markedly changed the flow near falling objects. Particle tracking (Fig. \ref{fig3}, inserts) revealed that fast shapes have long wakes with strong external recirculating and often transitioned to pitching behavior, consistent with \citet{ern2012wake} and \citet{willmarth1964steady}. In contrast, slower shapes had shorter wakes and only recirculation within the main wake. This effect is illustrated for  natural and mutated \emph{Amanchelier arborea} leaves in Video 1.
 
From these quantitative and qualitative data it is clear that symmetry, and to a lesser extent lobedness, is a strong predictor of high settling velocities. This lends additional support to the fast-leaf-hypothesis: symmetric forms settle fast, and no substantially faster shapes available to natural selection were identified. We acknowledge, however, that fitter shapes may exist among forms not based on random walk outlines \cite{hunt2007relative}, and that porosity (not considered here) could be an important flow-control parameter \cite{bozdag2023novo,cummins2018separated}.


\section{Discussion and Conclusion}
Our data demonstrate that settling aerodynamics strongly limits leaf morphology. Data from wildtype and as1-mutated leaves suggest that high symmetry is a critical factor in rapid settling (Fig. \ref{fig2}b). A comprehensive comparison to random forms confirms this picture and suggests that extant leaves are near-optimal (Fig. \ref{fig3}). At present, realistic shape mutants provide extant trees with minimal access to increased fitness (i.e., faster settling speed) by natural selection. However, mutated leaves which will occur due to climate change  \cite{peppe2011sensitivity,carvalho2021extinction,hodar2002leaf}, are generally inferior. This points toward a new positive feedback mechanism for atmospheric carbon.

\new{It is worthwhile to consider the limitations of this study in the context of non-deciduous plants. An important feature of non-deciduous leaves is that they are significantly denser than deciduous leaves. For instance, \citet{poorter2009causes} found that the leaf mass per area is up to 10 times larger (mean 2.5 times). This, of course, changes the settling dynamics because heavy objects are relatively less affected by drag and thus settle faster \cite{auguste2013falling}. While the relatively high density of non-deciduous leaves may have evolved primarily to mitigate effects of, for instance, drought, cold, and herbivory, nutrient recycling via rapid settling is another advantage. To compare to non-deciduous plants, we conducted settling experiments on leaf mimics variable area density, confirming previous studies of the positive correlation between speed and area density (Fig. \ref{FigS4}). }

With a renewed focus on the carbon cycle and forest health, it is more critical than ever to understand how carbon is managed by trees. The connection between leaf symmetry and the fate of leaves in forests and other ecosystems provides a simple, quantifiable link between the health of the ecosystem and the morphology of leaves. Climate change has been shown to heavily modulate tree leaf shapes  \cite{peppe2011sensitivity,carvalho2021extinction}, and can impact the symmetry of leaves \cite{hodar2002leaf}. The link between the canopy and the soil may also be a major factor in stopping outdated practices such as leaf litter harvesting  \cite{sayer2006leaflitternutrition,pote2004effects} that are still prevalent today. Moreover, patterns in the fossil record may have been influenced by leaf sedimentation \cite{boyce2002evolution}. Finally, we speculate that real and synthetic leaves may be a new and fruitful source of inspiration for aeronautics.

\bibliography{refs}

\appendix


\section{Appendices}
\subsection{Methods}
\subsubsection{Experimental Setup}
We have built a custom enclosure for laser cutting and observing the paper falling. This setup utilizes a roll of 2 ply blue workshop paper (Luca Eko Express, D-S, Denmark), which is pulled through a network of rollers to hold it flat while the laser is cutting. The enclosure is automated by a microcontroller (Arduino Leonardo, Arduino, Italy), which actuates the paper movement via a stepper motor, switch the lights, records the temperature, and switches the power for the laser cutter. We use a 5W diode laser cutter (Master 2S, Neje, China), which is held above the water tank (Nanotank 20, Amtra, Germany). This tank has dimensions of 25x25x30cm, with typical shapes having a maximum dimension of 2cm, cut close to the centre of the tank. Experiments are performed above ambient temperatures at $24.5\pm0.5\degree$C. To ensure rapid wetting of the paper, we use a solution of 0.1\% nitric acid (AR grade, Fisher, Denmark) and 10mL of detergent (Liquinox, Merck, Denmark) in $\sim 20$L of deionised water. This water is pumped into the tank warm ($\sim 40\degree$C) and allowed to cool to room temperature over several hours to minimize bubble formation. 
 
 \subsubsection{Optics and Image processing}
To observe the falling paper we use a Raspberry Pi HQ camera (Raspberry Pi Foundation, UK) with a 12mm fixed focal length lens (C11-1220-12M-P f12mm, Basler, Germany). This is attached to a Jetson Nano 4GB (Nvidia, USA), which uses a custom script based on FastMOT  \cite{yukai_yang_2020} to record the video and track the falling paper. The tracking is performed live with a CUDA accelerated background subtraction and a contour finding algorithm in the OpenCV package. All shapes are calibrated within a day by a disk in case of misalignments of the camera.

\subsubsection{Shape processing, quantification and generation}
All leaves originate from the LeafSnap dataset \cite{leafsnap_eccv2012} and the origin, symmetry and lobedness characteristics of each shape are summarized in the SM. The outlines of the leaves were extracted with OpenCV and the majority of petioles and deep corners were removed to prevent ignition of the paper. We then resize the shapes, space the points equally and dilate them to avoid artefacts from the 0.5mm burn width of the laser cutter. To avoid directionality bias of the paper, the shapes are randomly rotated before each cut, and the shape order is randomized within each experiment.

The assessment of the lobedness $L$ and symmetry $S$ were both performed similarly. Lobedness was calculated by fitting an oval of the same area to the assessed shape. This is done via a differential evolution algorithm that searches through the various oval centers, tilt angles, and aspect ratios to find an oval that minimizes the sum of the areas not encompassed by the oval-shape overlap or an XOR operation. Similarly, the symmetry value was calculated by a differential evolution algorithm that chose a mirror line through a given shape, where the shape was mirrored, and the non-overlapping areas of the shapes were minimized. The XOR operation was done numerically, translating the shapes into binary images (0 or 1), applying the operation bitwise across the image, and summing the image to determine the area. For a diagrammatic representation of the process, see Fig. \ref{Supps2}.

Brownian shapes were generated by making a simple schema where a point would be randomly added to a list that was one integer value up, down left or right from the previous integer. This would be repeated 500 times to generate a large shape. These points are then transformed into an image, the outer contour is found and then dilated to remove any sharp corners. For a diagrammatic representation of this process, please consult Fig. \ref{Supps3}.

\subsubsection{Particle flow field tracing}
To perform the flow field tracing, we generated a light sheet with a 100mW, $110\degree$ laser line module (MediaLas, Germany) positioned 10cm from the tank, and the tank was seeded by 20µm particles (Dynasphere, Norway) that were dispersed in 10\% detergent prior to mixing into the tank. For these experiments, we used identical optics as the shape tracking apparatus and an area scan camera (acA1920-155um, Basler, Germany) with a capture rate of 10fps. Videos of the objects falling are taken and images are extracted for analysis.
\subsection{Leaf shape data}
\begin{figure*}[]%
\centering
\includegraphics[width=1\textwidth]{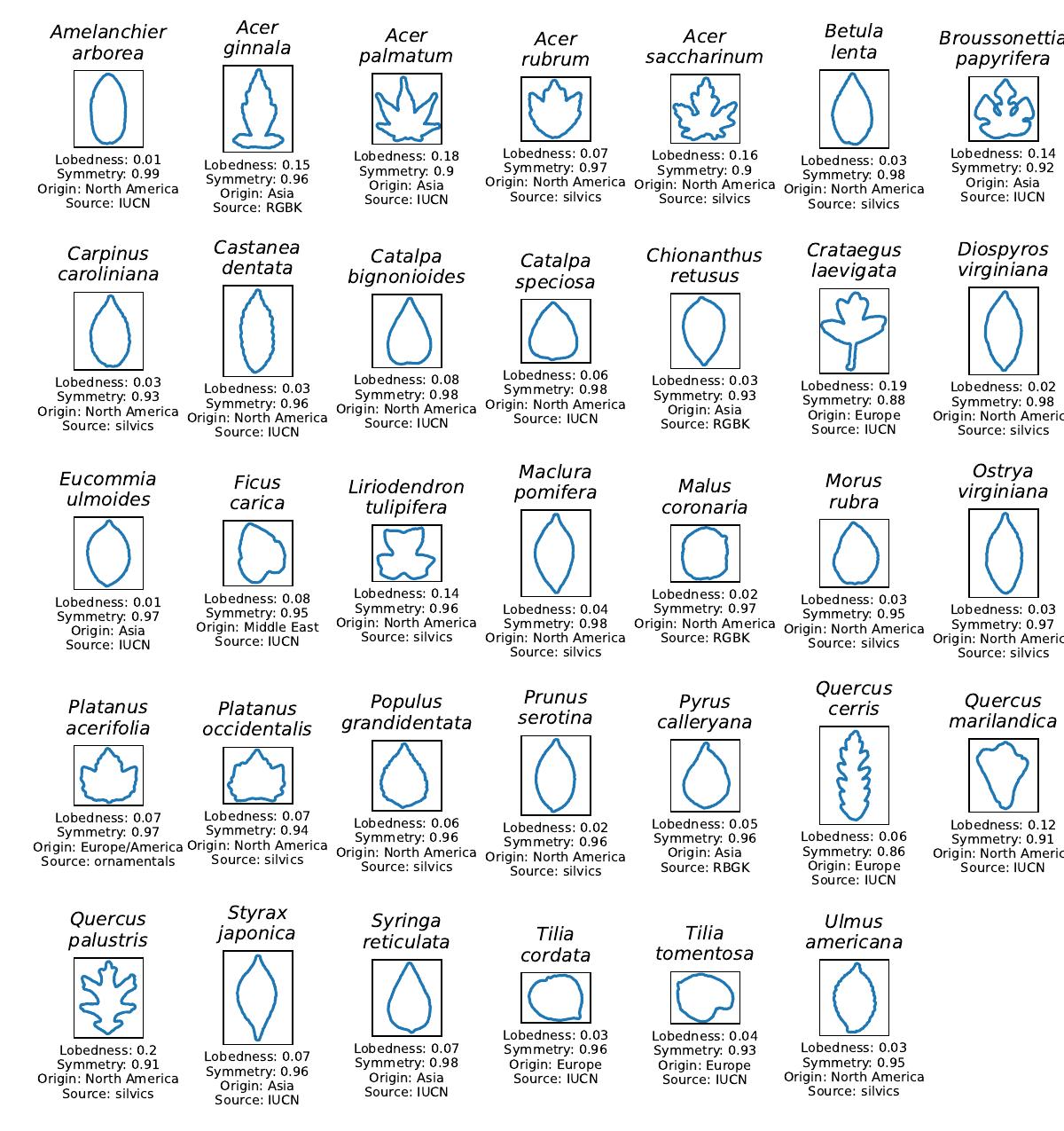}
\caption{Leaf shape data. Leaf outlines, lobedness and symmetry scores, as well as their region of origin. Sources IUCN \cite{IUCN}, RBGK \cite{RBGK}, silvics \cite{burns1990silvics}, and ornamentals \cite{li1996shade}, can be found in the bibliography.
}\label{Supps1}
\end{figure*}

The settling experiments (Figs. \ref{fig1}-\ref{fig3}) used the leaf forms shown in Fig. \ref{Supps1}. Detailed references are provided in the figure caption and in the bibliography.
\newpage
\clearpage

\subsection{Symmetry and Lobedness number}
\begin{figure}[h]%
\centering
\includegraphics[width=0.5\textwidth]{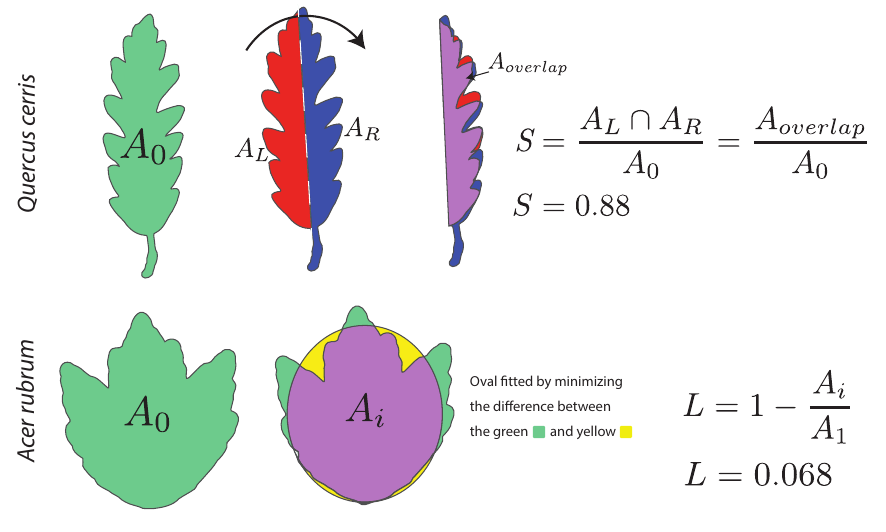}
\caption{A graphical explanation of the reflection symmetry ($S$,top) applied to an \textit{Quercus cerris} and lobedness ($L$, bottom) applied to an \textit{Acer rubrum} leaf. The reflection symmetry is found by choosing a cut line for each half of the leaf and minimizing the residual non-overlapping quantity. Symmetry $S$ is the overlapping quantity divided by the overall leaf area. The lobedness $L$ is found by minimizing the sum of the green and yellow areas via a simulated annealing algorithm. The area of the oval and the leaf are equal.
}\label{Supps2}
\end{figure}

The symmetry  $S$ and lobedness $L$ are defined in Fig. \ref{Supps2}.

\subsection{Random leaf shapes}
\begin{figure}[h]%
\centering
\includegraphics[width=0.5\textwidth]{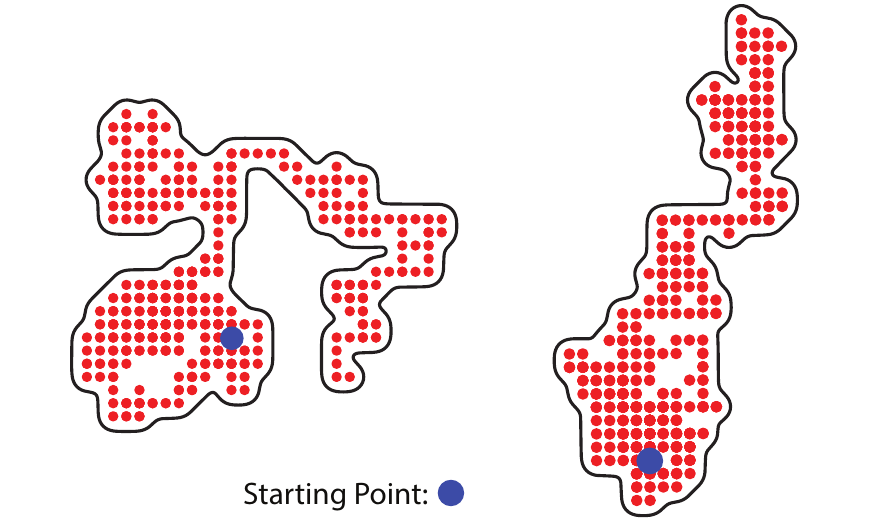}
\caption{Generating the Brownian shapes. Using simplified Brownian motion (assuming random movements in a grid) we generated a point set as seen in the figure. A shape was then generated from these points by finding the outline of the shape and then adding 1.5 particle diameters to the perimeter, creating a solid shape. 
}\label{Supps3}
\end{figure}

The procedure used to generate synthetic leaf outlines using a random walk is illustrated in Fig. \ref{Supps3}.

\subsection{Control experiment: Leaf area density}
\begin{figure}[h]%
\centering
\includegraphics[width=0.5\textwidth]{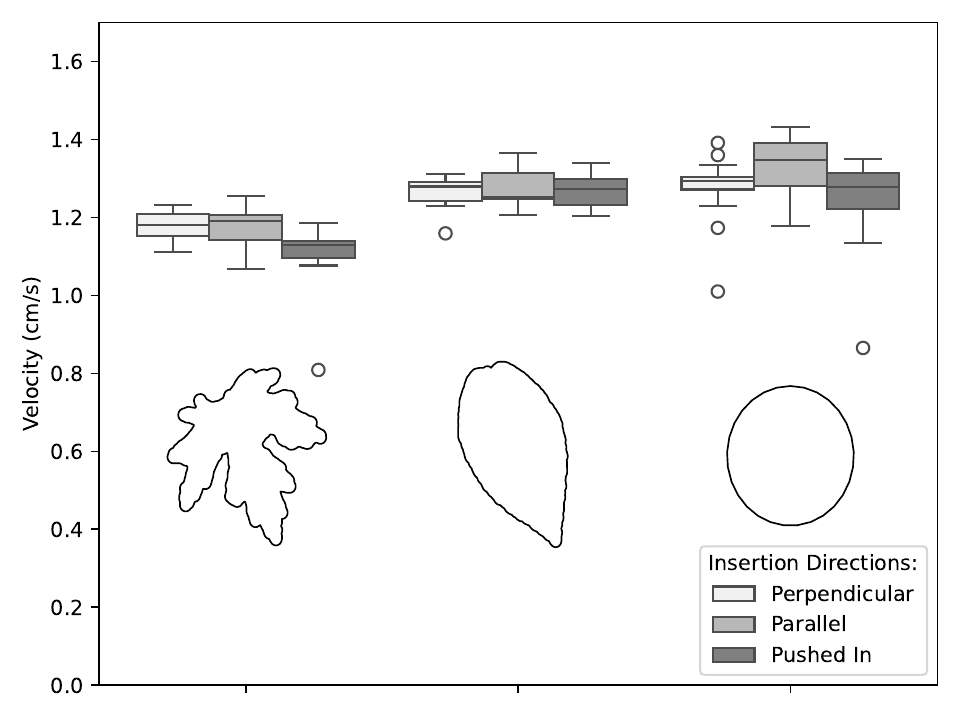}
\caption{Terminal settling velocity as a function of area density for a circular disk.}\label{FigS4}
\end{figure}

The effect of area density on settling is illustrated in Fig. \ref{FigS4}.

\subsection{Control experiment: Leaf area}
\begin{figure}[h]%
\centering
\includegraphics[width=0.5\textwidth]{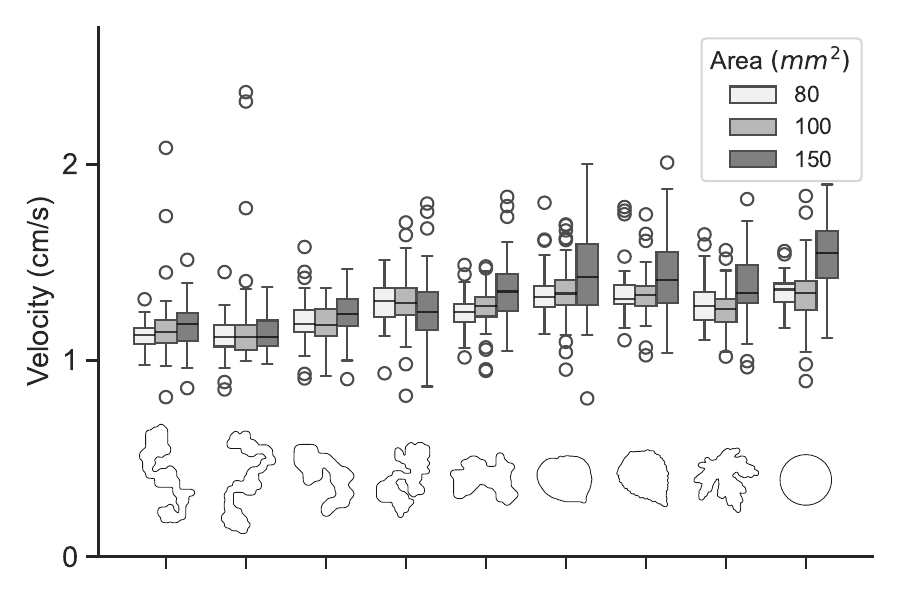}
\caption{Terminal settling velocity as a function of area for various leaf shapes.
}\label{FigS5}
\end{figure}

The effect of leaf area on settling is illustrated in Fig. \ref{FigS5}.

\subsection{Control experiment: Initial condition}
\begin{figure}[h]%
\centering
\includegraphics[width=0.45\textwidth]{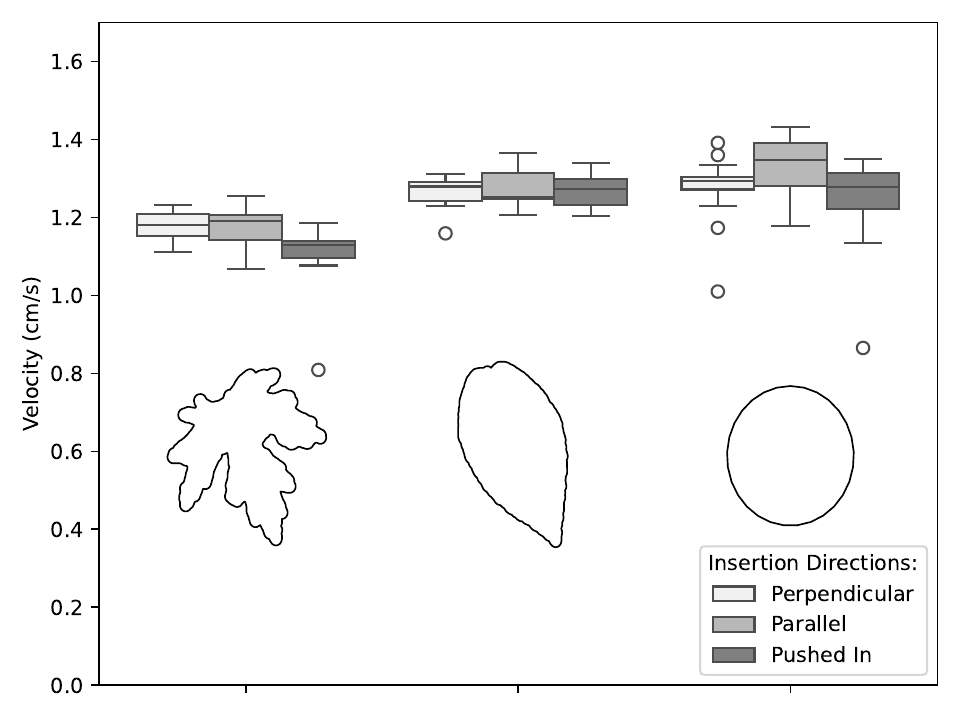}
\caption{Terminal settling velocity as a function initial condition.
}\label{FigS6}
\end{figure}

The effect of the initial condition at the air-water interface on settling is illustrated in Fig. \ref{FigS6}.

\subsection{Settling characteristics of asymmetric leaf mimics}
\begin{figure}[h]%
\centering
\includegraphics[width=0.2\textwidth]{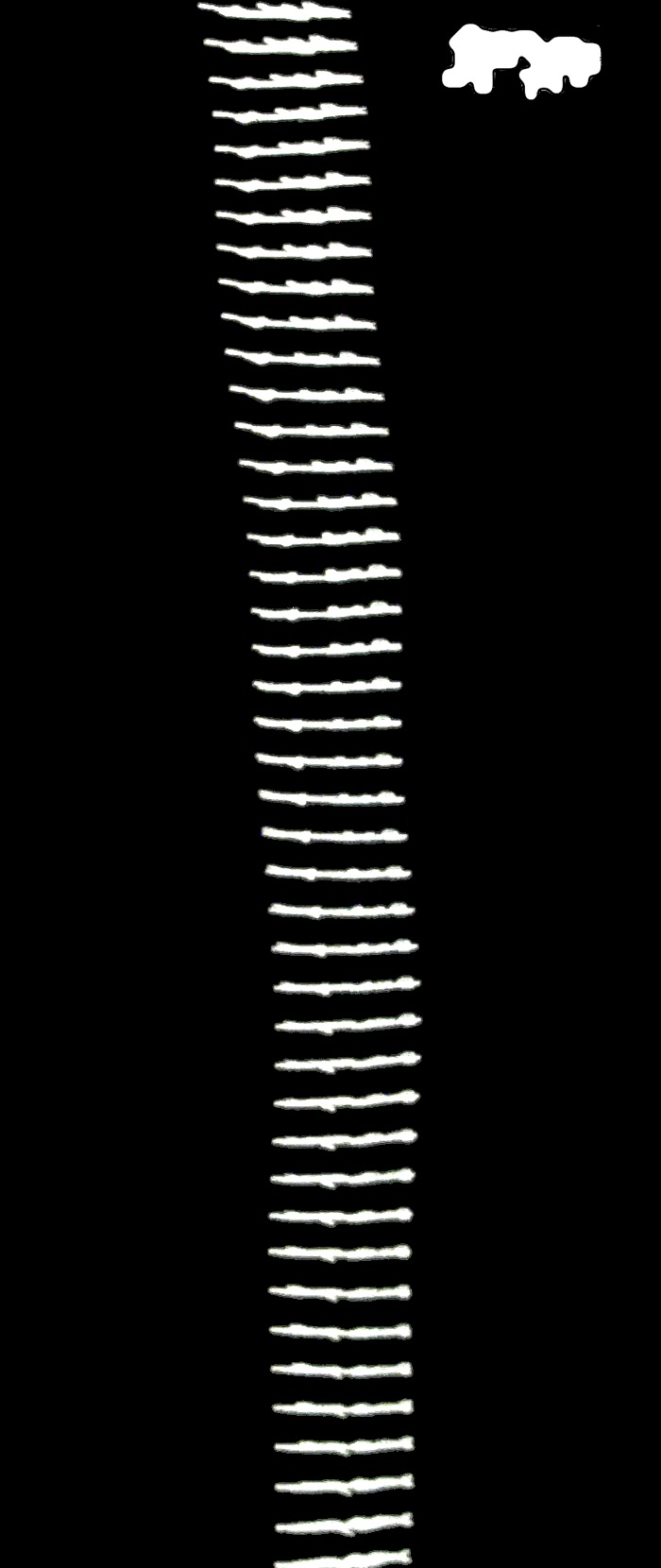}
\caption{Most asymmetric leaf mimics rotated slowly while falling steadily. The superposed images show an asymmetric leaf (area 100 mm$^2$, shape id: QbWYJ). The time between each frame is 0.25 s.
}\label{FigS7}
\end{figure}

Generally speaking, asymmetric leaf mimics rotated slowly while falling steadily (Fig. \ref{FigS7}). We speculate that irregular morphologies reduced flow features often associated with fluttering and tumbling.

\subsection{Magnified view of the flow patterns}
\begin{figure}[h]%
\centering
\includegraphics[width=0.5\textwidth]{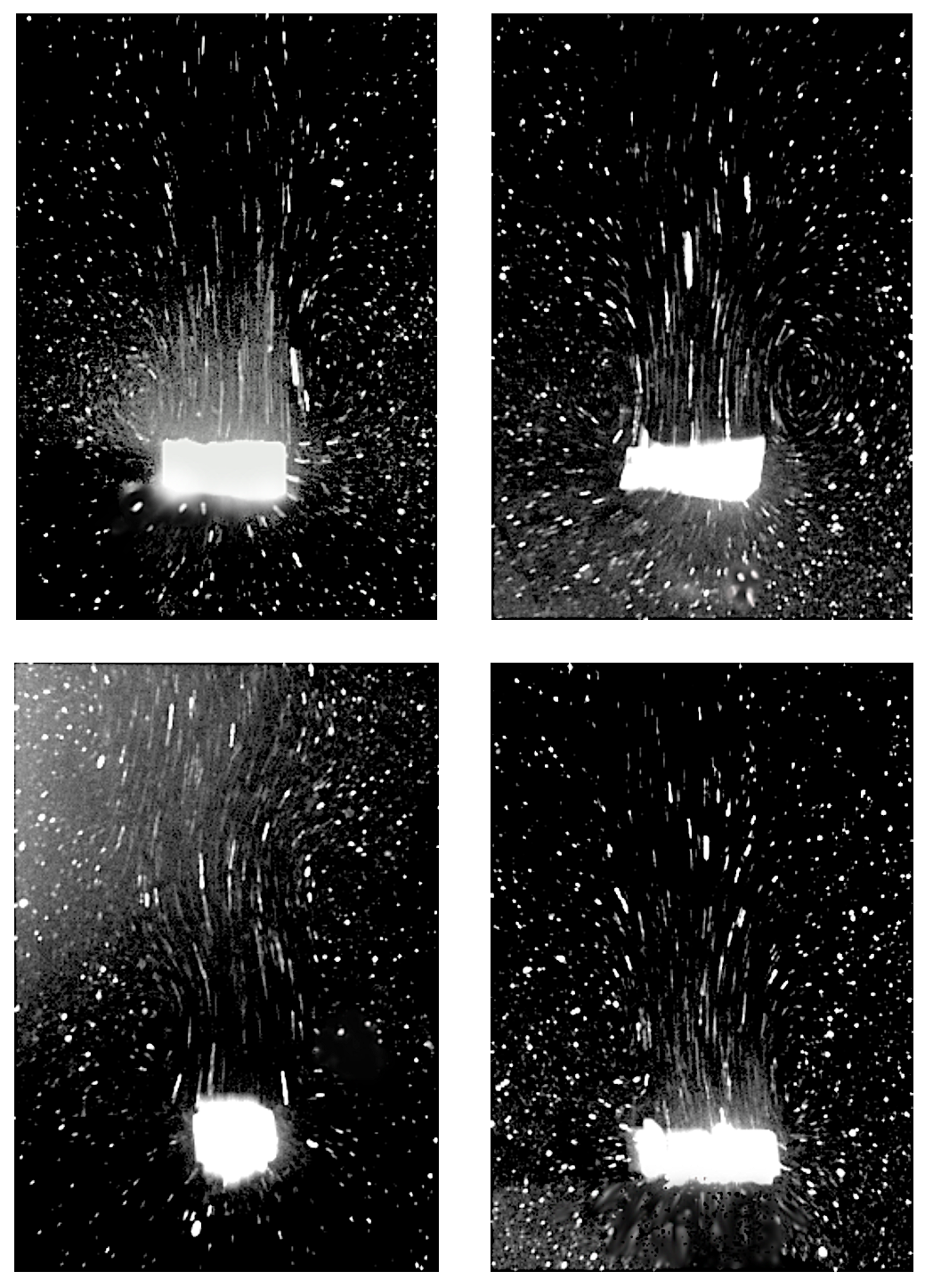}
\caption{Magnified view of the flow patterns shown in Fig. \ref{fig3}. Long exposure of a laser sheet and 0.02 mm tracer particles dispersed in the solution were used to generate the images.}\label{FigS8}
\end{figure}

\noindent A magnified view of the flow patterns shown in Fig. \ref{fig3} is provided in Fig. \ref{FigS8}.
\clearpage
\subsection{Settling characteristics as a function of the Reynolds number $Re$}
\begin{figure}[]%
\centering
\includegraphics[width=0.45\textwidth]{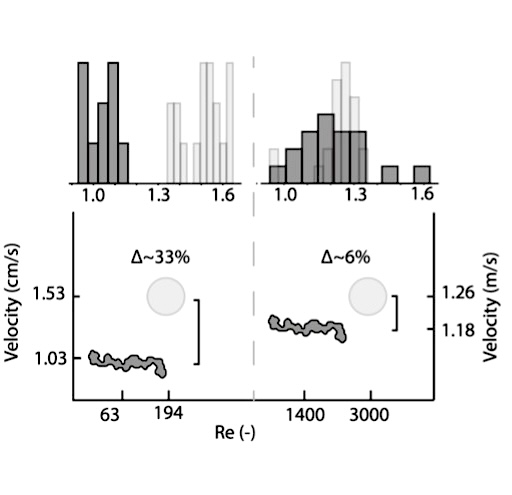}
\caption{The relative leaf mimic performance is sustained across $Re\approx 100-1000$. Settling speed depends on the leaf shape, and asymmetric leaves settle relatively slowly. Settling velocity plotted as a function of Reynolds number Re for the outlined shapes. The speed distributions are also shown (histograms). Re=63-174: wet laser cut shapes settling in water as described in the main paper. Re=1400-3000 data: dry laser cut leaf mimics settling in air. The dry samples were released using a remote-controlled electromagnetic clamp. To remove the intrinsic curvature from the rolled-up paper, the dry leaf mimics were gently folded along the central axis before release.}\label{FigS9}
\end{figure}

\noindent Experiments exploring settling characteristics as function of $Re$ are shown in \ref{FigS9}. The relative performance is sustained across $Re\approx 100-1000$, consistent with \citet{bagheri2016drag}.
\newpage
\subsection{Video 1}
\begin{figure}[ht!]%
\centering
\includegraphics[width=0.4\textwidth]{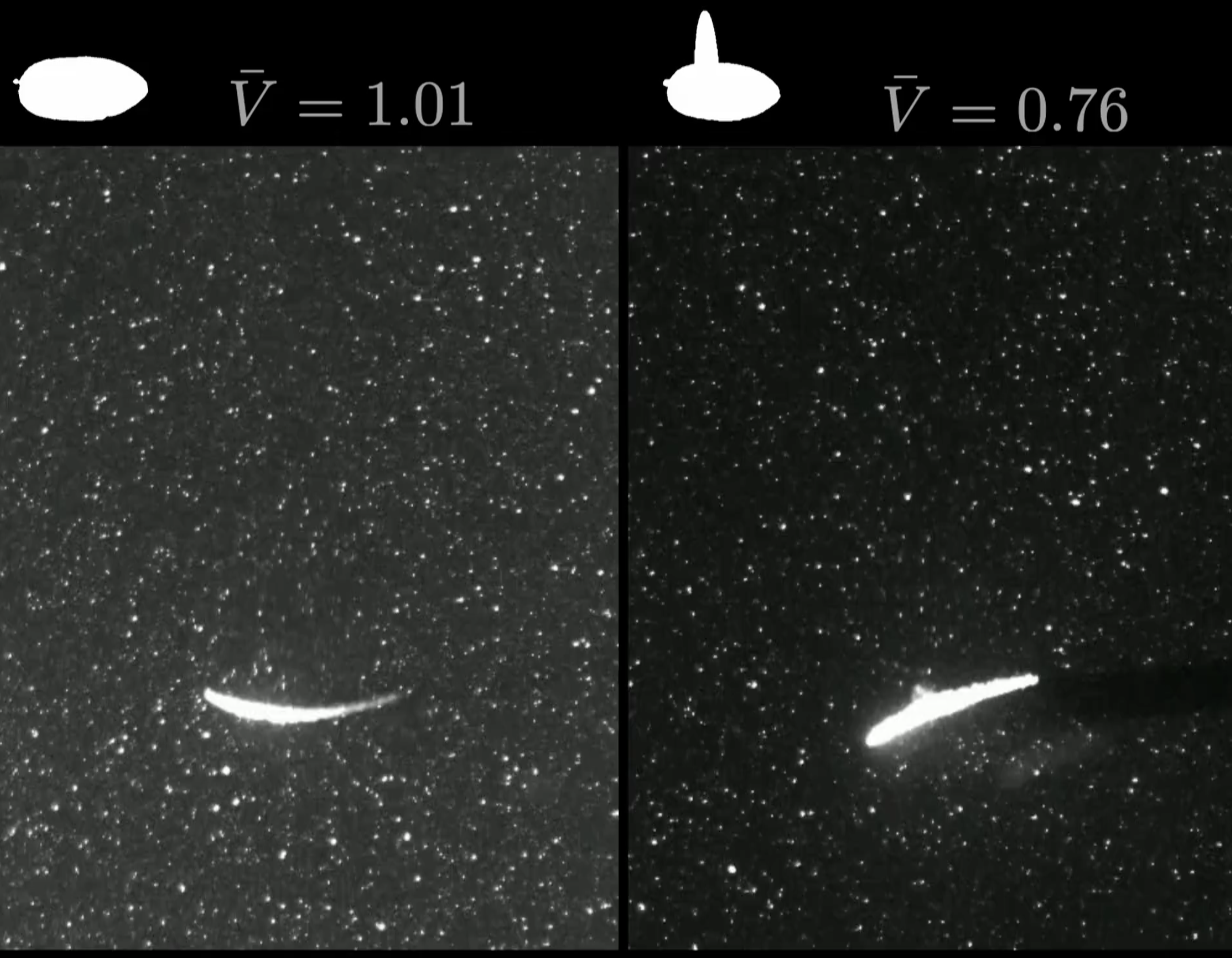}
\caption{Video 1: This video (available online) illustrates the settling of natural and mutated \emph{Amanchelier arborea} leaves. The wild-type (left, $\bar V = 1.01$) settles at nearly the same speed as a circular disk ($\bar V = 1$). However, the mutated leaf (right, $\bar V = 0.76$) is nearly $25\%$ slower. Particles suspended in the fluid medium illustrate the qualitative difference in flow patterns.}\label{Fig13}
\end{figure}
\end{document}